\documentclass{appolb}
\usepackage{graphics}
\usepackage{graphicx}
\usepackage{epsfig}
\usepackage{color}
\usepackage{hyperref}
\usepackage{latexsym}
\usepackage{amsmath}
\usepackage{amsthm}
\usepackage{amsfonts}
\usepackage{amssymb}
\usepackage{dsfont}
\usepackage{bm}
\usepackage{graphics,psfrag}
\usepackage{graphicx,psfrag}
\usepackage[tight]{units}

\newcommand{\FC}{\;,}

\newcommand{\RDH}[1]{{\mathrm{red}(#1)}}
\newcommand{\LMH}[1]{{\mathrm{lm}(#1)}}

\newcommand{\otf}{{\scriptstyle\frac{1}{2}}}
\newcommand{\ttf}{{\scriptstyle\frac{3}{2}}}

\usepackage{epsfig}

\begin{document}
\title{Confinement, chiral symmetry breaking and the mass generation of hadrons%
\thanks{Invited talk at Light Cone 2012, 8-13 July, 2012,
 Cracow, Poland}%
}
\author{L. Ya. Glozman
\address{Institute for Physics, Theoretical Physics Branch,
University of Graz, Universit\"atsplatz 5, A-8010 Graz, Austria}
}

\maketitle
\begin{abstract}

A key question to QCD is what mechanism generates the
hadron mass in the light quark sector, where both confinement
and chiral symmetry breaking are in the game. Are confinement and
chiral symmetry breaking in the vacuum uniquely interconnected? Can  
hadrons survive chiral symmetry restoration? If yes, what  happens
with their mass and what symmetries beyond the chiral symmetry are there? 
We review our recent insights. In particular, 
in a dynamical lattice simulation we artificially restore chiral
symmetry by removing the low-lying Dirac modes of the valence quark
propagators, which is a well defined procedure and keep gluodynamics intact. 
Hadrons survive this artificial chiral restoration and their mass is surprisingly
large. All hadrons fall into chiral multiplets and some of them
are degenerate, i.e. the spectrum reveals some higher symmetry, that includes the chiral 
symmetry as a subgroup. The $U(1)_A$ symmetry does not get restored after
removal of the chiral modes from the valence quarks.

\end{abstract}
  
\section{Introduction}

QCD is already 40 years old but we do not know yet the
answer to a key question about mass generation of hadrons.
In the light quark sector, with the quark masses of a few
MeV, practically the whole hadron mass consists of the 
energy of the quantized gluonic field. 
This straightforward answer from the trace anomaly of QCD
is correct but not satisfactory.
We are interested in the mechanism of the mass generation.
Are confinement and dynamical chiral symmetry breaking
interconnected and how do they contribute to the hadron mass? It is also
important to shed the light on this issue if we want to understand
the phase diagram of QCD. 

It was believed  that the chiral symmetry breaking
in the vacuum is the crucial phenomenon responsible for the
hadron mass generation: The hadron mass in the light quark
sector is determined mainly by the quark condensate of the
vacuum. This is certainly true for the pion, which as the (pseudo)
Goldstone boson originates from the dynamical chiral symmetry 
breaking.  Is it true, however,
that the nucleon,  the rho-meson and other hadron  masses also come mostly from
the quark condensate of the vacuum? Given this view, it was expected
that upon  chiral restoration masses of these and other hadrons should
drop off \cite{BR} and eventually beyond the chiral restoration phase transition (crossover)
the hadrons should dissappear. In other words, without the chiral symmetry breaking
in the vacuum there cannot be any confined hadrons.

The 't Hooft anomaly matching conditions \cite{HOOFT}  formally state, that
at zero temperature and density  in the confining mode
the chiral symmetry should be indeed spontaneously broken, though
they do not suggest any insight why it should be so. This generic
statement does not imply, however, that the hadron mass should be
made mostly of the quark condensate of the vacuum. The latter view had
essentially phenomenological \cite{SVZ,IOFFE} and model grounds starting from
the bag model in the past  up to contemporary Schwinger-Dyson approaches 
to hadrons. The 't Hooft anomaly matching conditions do not constrain,
however, the interrelation between the confinement and chiral symmetry
breaking at nonzero temperatures and densities.

Another argument, according to Casher \cite{Casher}, was that  the quark cannot
be confined without the chiral symmetry breaking in the vacuum. If so, hadrons cannot
exist in the world with unbroken chiral symmetry. However, it was shown that
the Casher argument is not general and can be easily bypassed \cite{G1}. In particular,
at least within the manifestly confining model hadrons with rather large
mass can still exist in a dense
medium at low temperatures where the chiral symmetry is restored \cite{GW}.

Another interesting issue is whether the highly excited hadrons reveal
or not effective restoration of the chiral symmetry \cite{G2}. If yes
(it should be confirmed or disproved experimentally),
then their mass  should not be influenced by the quark condensate of the
vacuum. There are also some experimental hints,  that hadrons in this regime reveal some
higher symmetry, that includes $SU(2)_L \times SU(2)_R$ as a subgroup.

We want to shed some light on all these questions. For this purpose
we can use lattice QCD as a tool to explore the interrelation between
confinement and chiral symmetry breaking and to check whether or not
hadrons can still exist in a world without breaking of  chiral symmetry
in a vacuum. If yes, what happens with their mass and what symmetries do they
have in this regime?

The idea, that the low-lying modes of the Dirac operator,
that are responsible for the chiral symmetry breaking are crucial
for hadron masses such as nucleon or $\rho$-meson, etc., has its roots 
in the instanton liquid model of the QCD vacuum \cite{INSTANTONS}. Subsequently,
the effect of the low-lying modes of the Dirac operator on  different
hadron correlators was studied on a small lattice within the quenched
approximation \cite{GRAND}. We pose just the opposite question: Will
hadrons survive if we remove the low-lying modes keeping the
gluodynamics intact? Such a procedure is a well-defined one \cite{COHEN}
and for
the $\pi$, $\rho$, $a_0$ and $a_1$ mesons was implemented in \cite{LS}.
Here we study both baryons and mesons as well their symmetries after
such artificial chiral restoration \cite{GLS}.

\section{The setup}

The quark condensate of the vacuum is related to  
a density of the lowest quasi-zero eigenmodes of the
Dirac operator \cite{BC}:

\begin{equation}\label{BC}
< 0 | \bar q q | 0 > = - \pi \rho(0).
\end{equation}

\noindent
Here first the
infinite volume limit is assumed at a finite quark  mass  and then the
chiral limit should be taken.

From the lattice calculations in a given finite volume we cannot
say a priori which and how many lowest eigenmodes of the Dirac operator are
responsible for the quark condensate of the vacuum. 
We  remove an increasing number of the lowest Dirac modes 
from the valence quark propagators, 

\begin{equation}\label{eq:red5}
S_\RDH{k}=S-S_\LMH{k}\equiv S- 
\sum_{i\le k} \mu_i^{-1} |{v_i}\rangle\langle{v_i}|\gamma_5\FC
\end{equation}

\noindent
and study the
effects of the (remaining) chiral symmetry breaking on the  masses of hadrons.
Here $S$ is the untrancated quark propagator,
the $\mu_i$ are the (real) eigenvalues of the Hermitian Dirac operator
$D_5 = \gamma_5 D$,
$|{v_i}\rangle$ are the corresponding eigenvectors and $k$ represents the
number of the removed lowest-lying modes. 

We perform our calculations on the unquenched two-flavor configurations
with chirally improved fermions \cite{GRAZ} on the lattice size of 2.4 fm at
the pion mass $m_\pi = 322$ MeV. 

\section{Existence of hadrons after unbreaking of chiral symmetry}

An interesting observation is that for all hadrons under our study,
except for a pion,  the quality of the exponential decay
of the correlators essentially improves by increasing the number
of removed eigenmodes. The exponential decay of the correlator with
the given quantum numbers indicates
that there is a  state with the same quantum numbers. Assume that
after removal of a sufficient amount of the low-lying
modes the exponential decay signals from all hadrons would
disappear. This would indicate that hadrons also disappear, i.e.
there is not confinement without the chiral modes of the Dirac operator.
We observe, however, a very clean signal from all hadrons, except for a
pion. The hadrons survive this artificial unbreaking of the chiral
symmetry. Even more, the nucleon and rho-meson masses do not decrease
upon chiral restoration, see Figs. 1 and 2!

\section{Meson degeneracyies and splittings and what they tell us}

If hadrons survive the restoration of the $SU(2)_L \times SU(2)_R$ chiral 
symmetry in the vacuum,  they must fall into parity-chiral multiplets
\cite{G2}. 
These multiplets  for the $J=1$ mesons are as follows:
\begin{center}
\begin{tabular}{lclcl}
$(0,0)      $&$:\quad$&$\omega(0,1^{--}) $&$\qquad$ &$f_1(0,1^{++}) $ \nonumber\vspace{6pt}\\
$(\otf,\otf)_a$&$:\quad$&$h_1(0,1^{+-})    $&$\qquad$ &$\rho(1,1^{--})$ \nonumber\vspace{6pt}\\
$(\otf,\otf)_b$&$:\quad$&$\omega (0,1^{--})$&$\qquad$ &$b_1(1,1^{+-}) $ \nonumber\vspace{6pt}\\
$(0,1)+(1,0)$&$:\quad$&$a_1(1,1^{++})    $&$\qquad$ &$\rho (1,1^{--}$)\nonumber
\end{tabular}
\end{center}

If the
$U(1)_A$ symmetry is unbroken
the states from two distinct multiplets $(\otf,\otf)_a$ and 
$(\otf,\otf)_b$ that have the same isospin but opposite spatial parity
are connected to each other by the $U(1)_A$ transformation.  In our
real world $U(1)_A$ is broken both  via the axial anomaly and
 via the quark condensate of the vacuum. In the world
with restored $SU(2)_L \times SU(2)_R \times U(1)_A$
symmetry a $\rho$ meson, that is the chiral
partner to the $h_1$ meson, must be degenerate with the $b_1$ state. 

On Fig. 1 we show the mass evolution of the isovector mesons
with $J=1$ upon the truncation of the low-lying modes. Both
the number of the removed modes $k$ as well as the maximal
energy $\sigma$ of these  modes are shown.

At the truncutaion energy $\sigma \sim 40$ MeV  the $\rho-a_1$ splitting
vanishes,  a direct indication of the chiral $SU(2)_L \times
SU(2)_R$  restoration in the physical states. The
large $b_1-\rho$ and $b_1-\rho'$ splittings persist, however. This 
means that the $U(1)_A$ breaking does not disappear. While
that $U(1)_A$ breaking component that is due to the chiral  condensate
should vanish with the condensate, the $U(1)_A$ breaking via the axial
anomaly still persists. Indeed, the quark determinant that contains
the $U(1)_A$ breaking in the vacuum, is not affected by our truncation of
the valence quarks. However, this result does show
that there is not direct
interconnection of the lowest lying modes of valence quarks and the
mechanism of the  $U(1)_A$ splittings in QCD. 

The $\rho$ and $\rho'$ mesons
become degenerate, too.  
These $\rho$ and $\rho'$ 
are different states because they appear in different eigenvalues of
the correlation matrix (i.e.
their eigenvectors are orthogonal) and because they are well split before the
removal of the low modes.  
This degeneracy indicates some higher symmetry
that includes chiral $SU(2)_L \times SU(2)_R$ as a subgroup. 

\begin{figure*}
\begin{center}
	\includegraphics[width=0.8\textwidth]{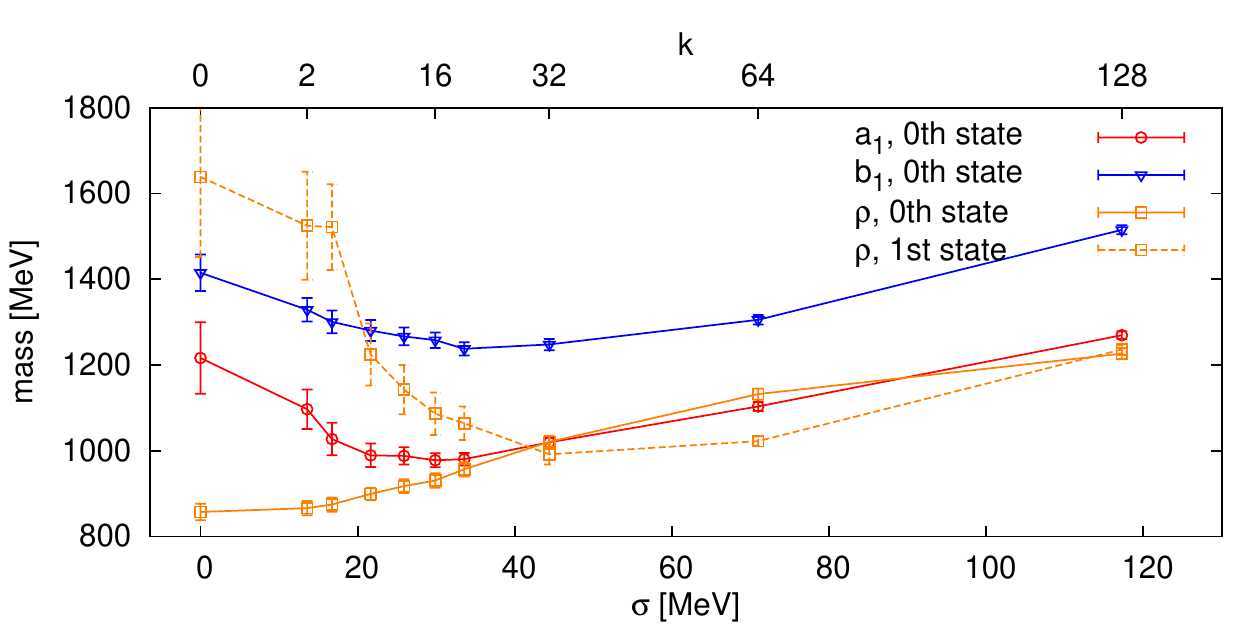}
	\hfill
	\caption{Meson masses
		as a function of the truncation level.}
        \label{mesons}
\end{center}
\end{figure*}

\begin{figure*}
\begin{center}	
	\includegraphics[width=0.8\textwidth]{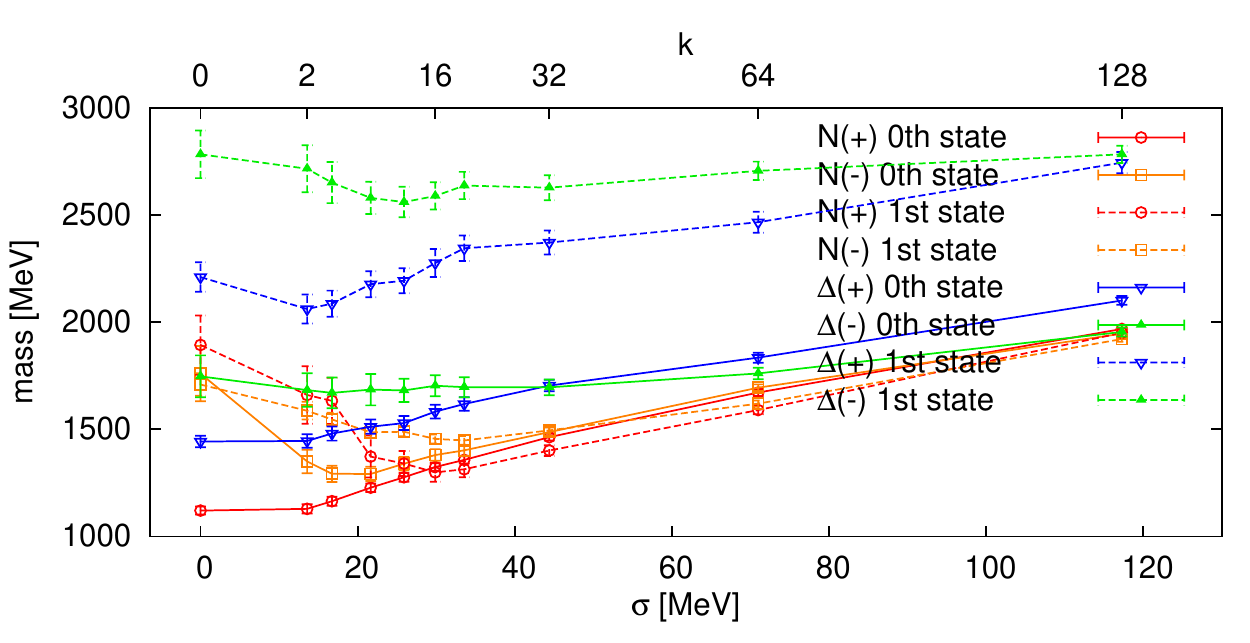}
	\hfill
	\caption{Baryon masses
		as a function of the truncation level.}
        \label{baryons} \end{center}
\end{figure*}

\section{Baryon chiral multiplets}

If baryons survive chiral restoration  they have to
fall into baryonic parity-chiral multiplets: 

\begin{equation}
(\otf,0) + (0,\otf)\,,~(\ttf,0) + (0,\ttf)\,,~ (\otf,1) + (1,\otf)\,. 
\label{bm}
\end{equation}

The first representation consists of nucleons of positive and negative
parity of the same spin. Another representation contains both
positive and negative parity $\Delta$'s with the same $J$. Finally, the
third representation combines one nucleon and one
Delta parity doublet with the same spin.

 We observe at least  two
degenerate nucleon parity doublets.  This  indicates
 a higher symmetry for the $J=I=\otf$ states.

\bigskip
{\bf Acknowledgments}

I am very grateful to Christian Lang and Mario Schr\"ock
for a fruitful collaboration on the present topics.
Support of the Austrian Science
Fund (FWF) through the grant P21970-N16 is acknowledged.

\end{document}